\documentclass{article}

\usepackage{graphicx}

\usepackage{amsmath}
\usepackage{amsfonts}
\usepackage{amssymb}
\def\abstract#1{\vskip 7mm 
        \begin{center}{\large Abstract}\par \smallskip
                \begin{minipage}[c]{12cm}
                        \small #1
                \end{minipage}
        \end{center}
}
\def\title#1{\begin{center}{\Large\bf #1}\end{center}}
\def\author#1{\vskip 5mm \begin{center}{#1}\end{center}}
\def\address#1{\begin{center}{\it #1}\end{center}}
\makeatletter
\@ifundefined{lesssim}{}{}
\@ifundefined{gtrsim}{}{}
\def\vereq#1#2{\lower3pt\vbox{\baselineskip1.5pt \lineskip1.5pt
\ialign{$\m@th#1\hfill##\hfil$\crcr#2\crcr\sim\crcr}}}
\makeatother

\begin{document}

\title{%
Ghost Condensation and Gravity in Higgs Phase
}
\author{%
Shinji Mukohyama\footnote{E-mail: mukoyama@phys.s.u-tokyo.ac.jp}
}
\address{%
Department of Physics, The University of Tokyo, Tokyo 113-0033, Japan}

\abstract{%
A tachyon is considered to be sick in the context of particle mechanics,
but in field theory just indicates instability of a background. We
consider a similar possibility that a ghost in field theory might be
just an indication of instability of a background and that it can
condense to form a different background around which there is no
ghost. We construct a low energy effective field theory based on the
derivative expansion around the stable background. Possible
applications are discussed, including dark energy, dark matter,
inflation and black hole. 
}

\section{Introduction}

Gravity at long distances shows us many interesting and mysterious
phenomena: flattening galaxy rotation curves, dimming supernovae, and so
on. These phenomena have been a strong motivation for the paradigm of
dark matter and dark energy, i.e. unknown components of the universe 
which show up only gravitationally. As we essentially do not know what
the dark matter and the dark energy are, however, it seems a healthy
attitude to consider the possibility that gravity at long distances
might be different from what we think we know.

This kind of consideration has been a motivation for attempts for IR
modification of gravity, e.g. massive gravity~\cite{Fierz-Pauli} and 
DGP brane model~\cite{DGP}. However, they are known to have a
macroscopic UV scale at around $1000$km, where effective field theories
break down~\cite{AGS,LPR}. This does not necessarily mean that these
theories cannot describe the real world, but implies that we need
non-trivial assumptions about the unknown UV completion. The recent
proposal of ghost condensation~\cite{paper1} evades at least this
problem and can be thought to be a step towards a consistent theory of
IR modification of general relativity.

In general, if we have scalar fields then there are many things we can 
play with them. In cosmology, inflation can be driven by the potential
part of a scalar field. It is also possible to drive inflation by the
kinetic part of a scalar field~\cite{k-inflation}. On the other hand,
scalar fields play important roles also in particle physics. A scalar
field is used for spontaneous symmetry breaking and to change force
laws in the Higgs mechanism. This is usually achieved by using a
potential whose global minimum is charged under the gauge symmetry. The
basic idea of ghost condensation is to break a symmetry and change a
force law by the kinetic part of a scalar field. In this sense the ghost
condensation can be considered as an analogue of Higgs mechanism.

\section{Tachyons and Ghosts}
\label{sec:tachyon-ghost}

A tachyon in particle mechanics is defined as a particle whose speed
exceeds the speed of light. It violates causality and, thus, a theory
with tachyons is considered to be sick in the context of particle
mechanics. However, this is not necessarily true in field theory. In 
field theory a tachyon is an excitation around a top of a
potential. In this case a tachyon just indicates an instability of the
background around which the theory is expanded. If the potential has
minima at other field values then the theory can be expanded around any
one of them and the low energy effective theory is healthy even in the
context of particle mechanics. One important point is that we cannot
talk about evolution from the unstable background with tachyons to the
stable background without tachyons within the context of particle 
mechanics. In order to describe such an evolution, we need a framework
more general than particle mechanics, namely field theory.

A ghost in field theory can be defined as a field with a wrong sign
kinetic term. This is equivalent to say that a ghost is a field with a 
negative norm. Because of this, a field theory with ghosts is thought to
be sick. Indeed, aside from the negative norm, the existence of a ghost
indicates instability since the energy associated with the ghost
excitation is not bounded from below at least perturbatively in the
context of field theory. This situation for ghosts in field theory is
somehow similar to that for tachyons in particle mechanics. Hence, it
seems natural to expect that a more general, perhaps not-yet-known
framework should be able to describe a ghost as just an indication of
instability of the background around which the theory is expanded. If
this is the case, the general framework should be able to describe the
dynamics from a background with ghosts to another background without
ghosts. What is important here is that, even without such a framework,
we can construct a low energy effective field theory (EFT) around the
latter background, which we call ghost condensate. For this reason, we
do assume the existence of the more general framework, or a UV
completion, but do not need to assume any properties of the UV
completion to describe low energy excitations of ghost condensate.

\section{Ghost Condensation}

The ghost condensation can be pedagogically explained by comparison with
the usual Higgs mechanism as in the table shown below. First, the order
parameter for ghost condensation is the vacuum expectation value (vev)
of the derivative $\partial_{\mu}\phi$ of a scalar field $\phi$, while
the order parameter for Higgs mechanism is the vev of a scalar field
$\Phi$ itself. Second, both have instabilities in their symmetric
phases: a tachyonic instability around $\Phi=0$ for Higgs mechanism and
a ghost instability around $\partial_{\mu}\phi=0$ for ghost
condensation. In both cases, because of the instabilities, the system
should deviate from the symmetric phase and the order parameter should
obtain a non-vanishing vev. Third, there are stable point where small 
fluctuations do not contain tachyons nor ghosts. For Higgs mechanism,
such a point is characterized by the vev of the order parameter
satisfying $V'=0$ and $V''>0$. On the other hand, for ghost condensation
a stable point is characterized by $P'=0$ and $P''>0$. Fourth, while the
usual Higgs mechanism breaks usual gauge symmetry and changes gauge
force law, the ghost condensation spontaneously breaks a part of Lorentz
symmetry (the time translation symmetry) and changes linearized gravity
force law even in Minkowski background. Finally, generated corrections
to the standard Gauss-law potential is Yukawa-type for Higgs mechanism
but oscillating for ghost condensation. 
\begin{center}
 \includegraphics[trim = 0 15 0 12 ,scale=0.4, clip]{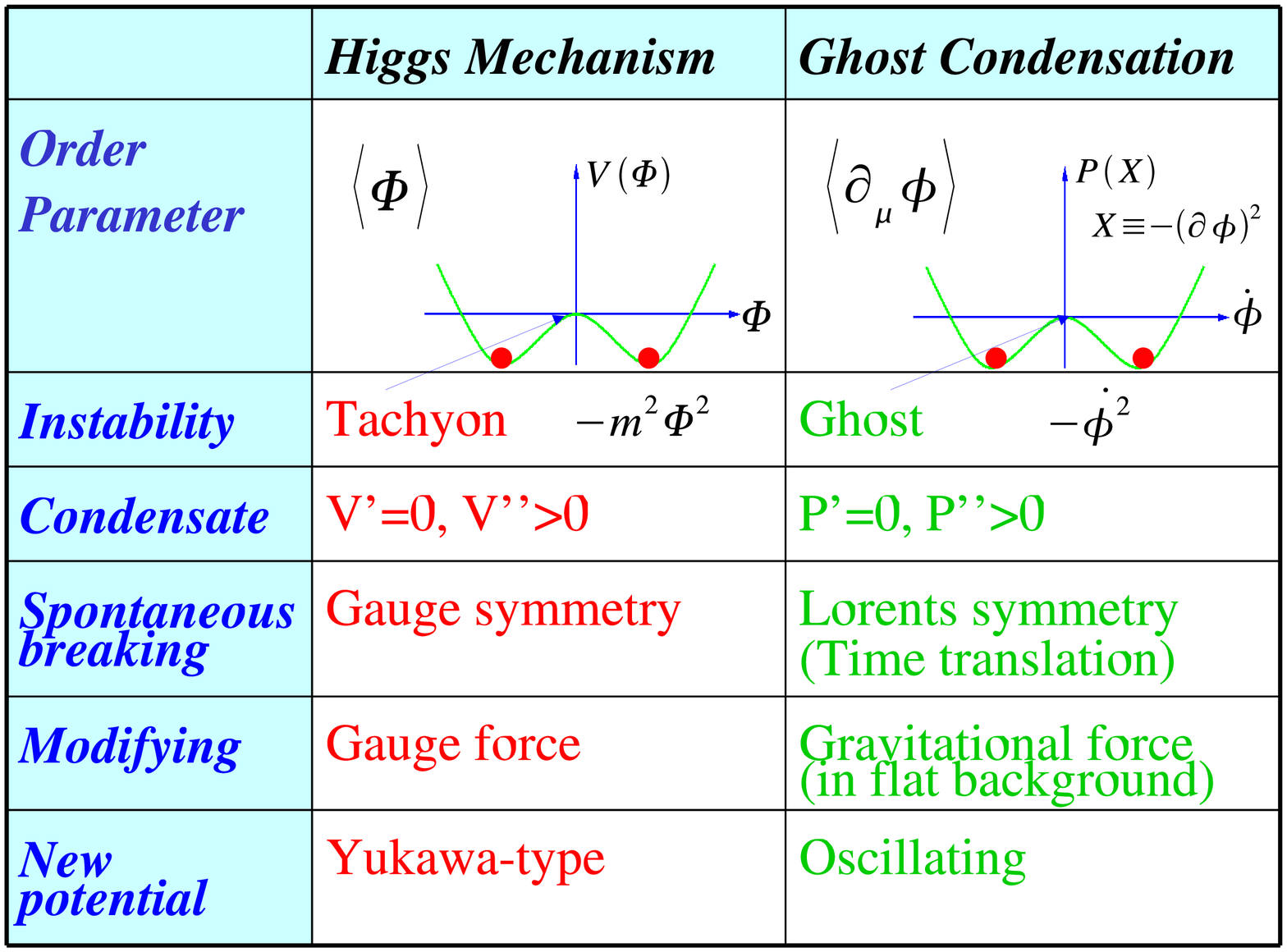}
\end{center}

For simplicity let us consider a Lagrangian 
$L_{\phi}=P(-(\partial\phi)^2)$ in the expanding FRW background with $P$ 
of the form shown in the upper right part of the table. We
assume the shift symmetry, the symmetry under the constant shift
$\phi\to\phi+c$ of the scalar field. This symmetry prevents potential 
terms of $\phi$ from being generated. The equation of motion for $\phi$
is simply $\partial_t[a^3P'\dot{\phi}]=0$, where $a$ is the scale factor
of the universe. This means that $a^3P'\dot{\phi}$ is constant and that
%============< EQUATION >==============%
%
\begin{equation}
 P'\dot{\phi} \propto a^{-3} \to 0 \quad (a\to\infty)
\end{equation}
%======================================%
as the universe expands. We have two choices: $P'=0$ or $\dot{\phi}=0$,
namely one of the two bottoms of the function $P$ or the top of the hill
between them. Obviously, we cannot take the latter choice since it is a
ghosty background and anyway unstable. Thus, we are automatically driven
to $P'=0$ by the expansion of the universe. In this sense the background
with $P'=0$ is an attractor. Thus, if there were inflation(s)
(irrespective of whether it is the usual potentially driven inflation,
k-inflation or ghost inflation) in the early universe then $P'$ is set
to an extremely small value.

Now let us consider a small fluctuation around the background with
$P'=0$. For $\phi=M^2 t + \pi$, the quadratic action for $\pi$ coming
from the Lagrangian $P$ is 
$\int d^4x[(P'(M^4)+M^4P''(M^4))\dot{\pi}^2-P'(M^4)(\nabla\pi)^2]$.
By setting $P'(M^4)=0$ we obtain the time kinetic term
$M^4P''(M^4)\dot{\pi}^2 $ with the correct sign. Unless the function $P$
is fine-tuned, $P''$ is non-zero at $P'=0$. This means that the
coefficient of the time kinetic term is non-vanishing and, thus, we do
not have the strong coupling issue which the massive gravity and the DGP
brane model are facing with. On the other hand, the coefficient of
$(\nabla\pi)^2$ vanishes at $P'=0$ and the simple Lagrangian $P$ does
not give us a spatial kinetic term for $\pi$. However, this does not
mean that there is no spatial kinetic term in the low energy EFT for
$\pi$. This just says that the leading spatial kinetic term is not
contained in $P$ and that we should look for the leading term in
different parts. Indeed, other terms like
$\tilde{P}((\partial\phi)^2)Q(\Box\phi)$ do contain spatial kinetic
terms for $\pi$ but the spatial-derivative expansion starts with the
fourth derivative: $(\nabla^2\pi)^2+\cdots$. If there is a non-vanishing
second-order spatial kinetic term $(\nabla\pi)^2$ then it can be
included in $P$ by redefinition and the redefined $P'$ goes to zero by
the expansion of the universe as shown above. Namely, the expansion of
the universe ensures that the spatial-derivative expansion starts from 
$(\nabla^2\pi)^2+\cdots$. Combining this spatial kinetic term with the
previously obtained time kinetic term and properly normalizing $\pi$, we
obtain the low energy effective action of the form
%============< EQUATION >==============%
%
\begin{equation}
 \int d^4x\left[ \frac{1}{2}\dot{\pi}^2
	   - \frac{\alpha}{M^2}(\nabla^2\pi)^2 + \cdots \right],
 \label{eqn:effective-action}
\end{equation}
%======================================%
where $\alpha$ is a dimensionless parameter of order unity. One might
worry that other (nonlinear) terms in effective theory such as
$\dot{\pi}(\nabla\pi)^2$ might mess up the effective action. In fact,
it turns out that all such terms are irrelevant at low
energy~\cite{paper1}. An important fact to show this is that the scaling
dimension of $\pi$ is not the same as its mass dimension $1$ but is
$1/4$, reflecting the situation that the Lorentz symmetry is broken
spontaneously. Moreover, it is also straightforward to show that all
spurious modes associates with higher time derivative terms such as
$(\ddot{\phi})^2$ have frequency above the cutoff $M$ and, thus, should
be ignored. In this sense, we are assuming the existence of a UV
completion but not assuming any properties of it. Finally, it must be
noted that the effective action of the form (\ref{eqn:effective-action})
is stable against radiative corrections. Indeed, the only would-be
more relevant term in the effective theory is the usual spatial kinetic
term $(\nabla\pi)^2$, but its coefficient $P'$ is driven to an extremely
small value by the expansion of the universe even if it is radiatively 
generated.

The effective action (\ref{eqn:effective-action}) would imply the low
energy dispersion relation for $\pi$ is 
$\omega^2\simeq\alpha k^4/M^2$. However, since the background
spontaneously breaks Lorentz invariance, $\pi$ couples to gravity in the
linearized level even in Minkowski or de Sitter  background. Hence, 
mixing with gravity introduces an order $M^2/M_{pl}^2$ correction to the
dispersion relation. As a result the dispersion relation in the presence
of gravity is $\omega^2\simeq\alpha k^4/M^2-\alpha M^2k^2/2M_{pl}^2$.
This dispersion relation leads to IR modification of gravity due to
Jean's instability. Note that there is no ghost around the stable
background $P'=0$ and the Jeans's instability is nothing to do with a
ghost.

As stated in the end of Sec.~\ref{sec:tachyon-ghost}, we actually do not
need to specify a concrete way of the spontaneous symmetry breaking in
order to construct the EFT around the stable background. In this sense,
the ghost around $\dot{\phi}=0$ has nothing to do with the construction
of the EFT around $P'=0$. Indeed, it is suffice to assume the symmetry
breaking pattern, namely from the full $4D$ Lorentz symmetry to the $3D$
spatial diffeomorphism~\cite{paper1}.

Note that the ghost condensate provides the second most symmetric class
of backgrounds for the system of field theory plus gravity. The most 
symmetric class is of course maximally symmetric solutions: Minkowski,
de Sitter and anti-de Sitter. The ghost condensate minimally breaks the
maximal symmetry and introduces only one Nambu-Goldstone boson.

\section{Possible Applications}

{\bf Dark energy:}
In the usual Higgs mechanism, the cosmological constant (cc) would be
negative in the broken phase if it is zero in the symmetric
phase. Therefore, it seems difficult to imagine how the Higgs mechanism
provides a source of dark energy. On the other hand, the situation is
opposite with the ghost condensation: the cc would be positive in the
broken phase if it is zero in the symmetric phase. Hence, while this by
itself does not solve the cc problem, this can be a source of dark
energy.

{\bf Dark matter:}
If we consider a small, positive deviation of $P'$ from zero then the
homogeneous part of the energy density is proportional to $a^{-3}$ and
behaves like dark matter. Inhomogeneous linear perturbations around the
homogeneous deviation also behaves like dark matter. However, at this
moment it is not clear whether we can replace dark matter with ghost
condensate. We need to see if it clumps properly. Ref.~\cite{paper3} can
be thought to be a step towards this direction.

{\bf Inflation:}
We can also consider inflation within the regime of the validity of the
EFT with ghost condensation. In the very early universe where $H$ is
higher than the cutoff $M$, we do not have a good EFT describing the
sector of ghost condensation. However, the contribution of this sector
to the total energy density $\rho_{tot}$ is naturally expected to be
negligible: $\rho_{ghost}\sim M^4 \ll M_p^2H^2 \simeq \rho_{tot}$. As 
the Hubble expansion rate decreases, the sector of ghost condensation
enters the regime of validity of the EFT and the Hubble friction drives
$P'$ to zero. If we take into account quantum fluctuations then $P'$ is
not quite zero but is 
$\sim (H/M)^{5/2}\sim (\delta\rho/\rho)^2\sim 10^{-10}$ in the end of
ghost inflation. In this way, we have a consistent story, starting from 
the outside the regime of validity of the EFT and dynamically entering
the regime of validity. All predictions of the ghost inflation are
derived within the validity of the EFT, including the relatively low-$H$
de Sitter phase, the scale invariant spectrum and the large
non-Gaussianity~\cite{paper2}.

{\bf Black hole:}
In ref.~\cite{paper4} we consider the question ``what happens near a
black hole?'' A ghost condensate defines a hypersurface-orthogonal
congruence of timelike curves, each of which has the tangent vector
$u^{\mu}=-g^{\mu\nu}\partial_{\nu}\phi$. It is argued that the ghost
condensate in this picture approximately corresponds to a congruence of
geodesics and the accretion rate of the ghost condensate into a black
hole should be negligible for a sufficiently large black hole. This
argument is confirmed by a detailed calculation based on the
perturbative expansion w.r.t. the higher spatial kinetic term. The
essential reason for the smallness of the accretion rate is the same as
that for the smallness of the tidal force acted on an extended object
freely falling into a large black hole.

\section*{Acknowledgements}

The author would like to thank Nima Arkani-Hamed, Hsin-Chia Cheng, Paolo
Creminell, Markus Luty, Jesse Thaler, Toby Wiseman and Matias
Zaldarriaga for collaboration on the subject.

\end{document}